\documentclass[twocolumn,trackchanges]{aastex631}
\usepackage{natbib}
\usepackage{epsfig}
\usepackage{graphicx}
\usepackage{subfigure}
\usepackage{float}
\usepackage{amsmath}
\usepackage{color}
\usepackage{amssymb}
\usepackage{amsfonts}
\usepackage{apjfonts}

\begin{document}

\title{A Comparative Analysis of Scale-invariant Phenomena in Repeating Fast Radio Bursts and Glitching Pulsars}

\correspondingauthor{Jun-Jie Wei}
\email{jjwei@pmo.ac.cn}

\author{Chong-Yu Gao}
\affiliation{Purple Mountain Observatory, Chinese Academy of Sciences, Nanjing 210023, China}
\affiliation{School of Astronomy and Space Sciences, University of Science and Technology of China, Hefei 230026, China}

\author[0000-0003-0162-2488]{Jun-Jie Wei}
\affiliation{Purple Mountain Observatory, Chinese Academy of Sciences, Nanjing 210023, China}
\affiliation{School of Astronomy and Space Sciences, University of Science and Technology of China, Hefei 230026, China}

\begin{abstract}
The recent discoveries of a remarkable glitch/antiglitch accompanied by fast radio burst (FRB)-like bursts
from the Galactic magnetar SGR J1935+2154 have revealed the physical connection between the two. In this work,
we study the statistical properties of radio bursts from the hyperactive repeating source FRB 20201124A and
of glitches from the pulsar PSR B1737--30. For FRB 20201124A, we confirm that the probability density functions of
fluctuations of energy, peak flux, duration, and waiting time well follow the Tsallis $q$-Gaussian distribution.
The derived $q$ values from $q$-Gaussian distribution keep approximately steady for different temporal
interval scales, which indicate that there is a common scale-invariant structure in repeating FRBs.
Similar scale-invariant property can be found in PSR B1737--30's glitches, implying an underlying association
between the origins of repeating FRBs and pulsar glitches. These statistical features can be well understood
within the same physical framework of self-organized criticality systems.
\end{abstract}
%Given the fact that the high-energy bursts from FRB 20201124A have a much higher burst rate than expected
%based on the energy distribution of lower-energy bursts, we separately investigate the scale-invariant behaviors
%of low- and high-energy bursts of FRB 20201124A. We found that the $q$ values of low- and high-energy bursts
%are different, which


\section{Introduction}
\label{sec:intro}

Fast radio bursts (FRBs) are intense millisecond-duration astronomical transients with mysterious physical origins
\citep{lorimer2007,2021SCPMA..6449501X,2022A&ARv..30....2P,2023RvMP...95c5005Z}. Some FRB sources,
such as FRB 20121102 \citep{2016Natur.531..202S} and FRB 20201124A \citep{2022ApJ...927...59L}, have been seen to
burst repeatedly, but it is still unclear whether repeating FRBs are prevalent or rare sources. Nonetheless,
the statistical analysis of the repetition pattern may shed new light on the physical nature and emission mechanisms
of FRBs (e.g., \citealt{chime2020,2020rajwade,2020Natur.587...45Z,2021cruces,2021Natur.598..267L}).

By studying the statistical properties of the repeating bursts from FRB 20121102, \cite{Lin2020} and \cite{Wei2021}
found that the probability density functions (PDFs) of fluctuations of energy, peak flux, and duration well follow
a $q$-Gaussian distribution. Here the fluctuations are defined as the size (energy, peak flux, or duration) differences
at different times. Furthermore, such a $q$-Gaussian behavior does not depend on the time interval adopted for the fluctuation,
i.e., the $q$ values in $q$-Gaussian distributions are approximately equal for different temporal interval scales,
which implies that there is a scale-invariant structure in FRB 20121102 \citep{Lin2020,Wei2021}. Analogous scale-invariant
characteristics have also been found in earthquakes \citep{Caruso2007,Wang2015}, soft gamma-ray repeaters (SGRs;
\citealt{Chang2017,Wei2021,2022MNRAS.510.1801S}), X-ray flares \citep{Wei2023} and precursors \citep{2023ApJ...955L..34L}
of gamma-ray bursts (GRBs), and gamma-ray flares of the Sun and the blazar 3C 454.3 \citep{2023ApJ...959..109P}. 
It is suggested that scale invariance is the most essential hallmark of a self-organized criticality (SOC) system
\citep{Caruso2007,Wang2015}. A generalized definition of SOC is that a nonlinear dissipative system with external 
drives input will self-organize to a critical state, at which a small local disturbance would generate an 
avalanche-like chain reaction of any size within the system \citep{bak1987}. Since the concept of SOC was proposed
\citep{Katz1986,bak1987}, it has been extensively used to explain the dynamical behaviors of astrophysical phenomena
\citep{2011soca.book.....A,2012A&A...539A...2A,2014ApJ...782...54A,2015ApJ...814...19A,2016SSRv..198...47A}.

On 2020 April 28, a bright radio burst FRB 20200428 was independently detected by CHIME \citep{2020Natur.587...54C}
and STARE2 \citep{2020Natur.587...59B} in association with a hard X-ray burst from a Galactic magnetar named SGR J1935+2154
\citep{2020ApJ...898L..29M,2021NatAs...5..378L,2021NatAs...5..372R,2021NatAs...5..401T}. This discovery provided
the first evidence for the magnetar origin of at least some FRBs \citep{2020Natur.587...45Z}. Since this event,
\emph{NICER} and \emph{XMM-Newton} telescopes have been monitoring SGR J1935+2154 regularly in the 1--3 keV band.
A phase-coherent timing analysis of X-ray pulses from the source was employed and a large spin-down glitch (also
referred to as ``antiglitch'') with $\Delta \nu =1.8^{+0.7}_{-0.5}$ $\mu$Hz was detected on 2020 October 5 ($\pm1$ day)
\citep{2023NatAs...7..339Y}. Subsequently, three FRB-like radio bursts emitted from SGR J1935+2154 were detected by
the CHIME/FRB system on 2020 October 8 \citep{2020ATel14074....1G}. More surprisingly, \cite{2022arXiv221103246G} utilized
the observations from \emph{NICER}, \emph{NuSTAR}, \emph{Chandra}, and \emph{XMM-Newton} to study the timing properties
of SGR J1935+2154 around the epoch of FRB 20200428, and reported that a giant spin-up glitch with $\Delta \nu =19.8\pm1.4$
$\mu$Hz occurred approximately $3.1\pm2.5$ day before FRB 20200428.

Glitches have now been observed in over a hundred pulsars, which are characterized as sudden increases 
in the rotational frequency of the star.
Despite several decades of research, the physical mechanisms of glitches are still not completely understood, but probably
involve an interaction between the neutron star's outer elastic crust and the superfluid component that lies within
(see \citealt{2015IJMPD..2430008H,2022RPPh...85l6901A} for reviews). The spin-down events, known as antiglitches,
are even rarer than glitches, and lack of observational data. The exact origin of antiglitches is also still open
to debate \citep{2000ApJ...543..340T,2014ApJ...782L..20H,2015MNRAS.453..522M}. The temporal coincidences between
the glitch/antiglitch and FRB-like bursts of SGR J1935+2154 reveal the physical connection between the two
\citep{2022arXiv221103246G,2023MNRAS.523.2732W,2023NatAs...7..339Y}. Motivated by this, we here examine the scale-invariant
similarities between repeating FRBs and glitching pulsars.

Since the sample size of most repeaters in the past was small, FRB 20121102 with relatively 
more repeating bursts is so far the only source that has been used to study the scale-invariant property
\citep{Lin2020,Wei2021}. As more hyperactive repeaters have been detected, it is interesting to find out 
whether other FRBs share the same scale-invariant property.
One of the most hyperactive repeating sources to date is FRB 20201124A, with 2744 independent bursts detected by
the Five-hundred-meter Aperture Spherical radio Telescope (FAST), representing the largest sample for a single
observation \citep{Xu2022,2022RAA....22l4002Z}. In addition, \cite{Kirsten2023} reported the detection of 46
high-energy bursts\footnote{Note that in this paper, ``high-energy burst'' refers to bursts
with large values of their isotropic energies, rather than bursts detected at higher observing frequencies.}
from FRB 20201124A in more than 2000 hours using four small 25--32-m class radio telescopes.
The energy distribution slope of these high-energy bursts is much flatter than that of the high-energy tail
in the FAST data, suggesting that ultra-high-energy ($E>3\times 10^{39}$ erg) bursts occur at a much higher rate
than expected based on FAST observations of lower-energy bursts \citep{Kirsten2023}. The study on the scale-invariant
properties of low- and high-energy bursts would be helpful for judging whether the highest-energy bursts originate from
a different emission mechanism or emission region at the progenitor source. On the other hand, for glitching pulsars,
their scale-invariant property has not yet been explored. However, because glitches are infrequent events,
the number of detected glitches in most of pulsars is not substantial enough to carry out robust statistical analyses
on individual bases \citep{2019A&A...630A.115F}. To date, 37 glitches in PSR B1737--30 have been detected, 
and these glitches exhibit a power-law size distribution. It should be underlined that the SOC systems 
would be slowly driven toward a critical state of the instability threshold of the entire system, leading to 
scale-free power-law size distributions \citep{2014ApJ...782...54A}. Thus, the emergence of a power-law size 
distribution is another fundamental property that SOC systems have in common
\citep{2011soca.book.....A,2012A&A...539A...2A,2015ApJ...814...19A}, suggesting that PSR B1737--30 has enough 
glitches for studying its scale-invariant feature.

In this work, we investigate and examine the similar scale-invariant phenomena between FRB 20201124A 
and glitches of PSR B1737--30. Moreover, in view of the fact that the burst energy distribution of FRB 20201124A
flattens towards the highest-observed energies, the sensitivity limits of FAST and 25--32-m class radio telescopes 
are different, and the statistical characteristics of 46 high-energy bursts would be submerged in the large amount 
of FAST data for the combined analysis, the scale-invariant behaviors of low- and high-energy bursts from FRB 20201124A
are thus studied separately. The rest parts of this article are organized as follows. In Section \ref{sec:FRB}, 
we make the scale-invariant analyses on energy, peak flux (luminosity), duration, and waiting time of low- and 
high-energy bursts from FRB 20201124A, respectively.
In Section \ref{sec:glitch}, taking PSR B1737--30 as an example, the distributions of fluctuations of glitch size
and of waiting time between consecutive glitches are presented for the first time. Finally, relevant conclusions
are drawn in Section \ref{sec:sum}.

\section{Scale Invariance in FRB 20201124A}
\label{sec:FRB}

\begin{figure*}
\begin{center}
\vskip-0.1in
\includegraphics[width=0.45\textwidth]{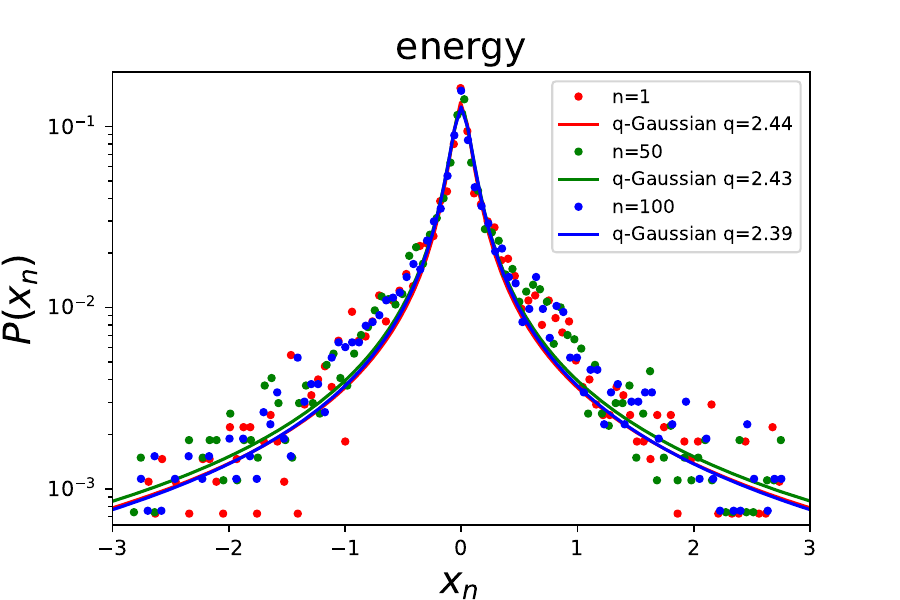}
\includegraphics[width=0.45\textwidth]{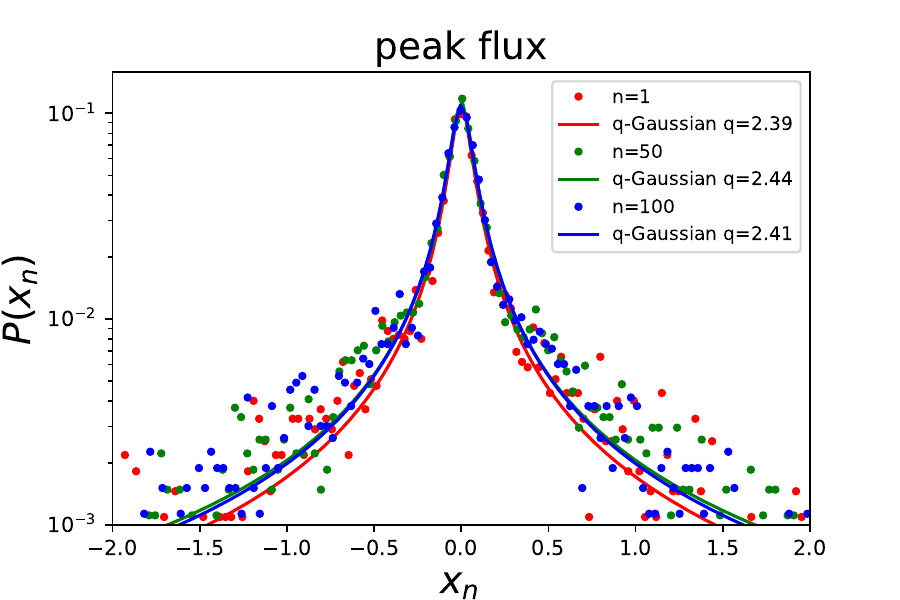}
\includegraphics[width=0.45\textwidth]{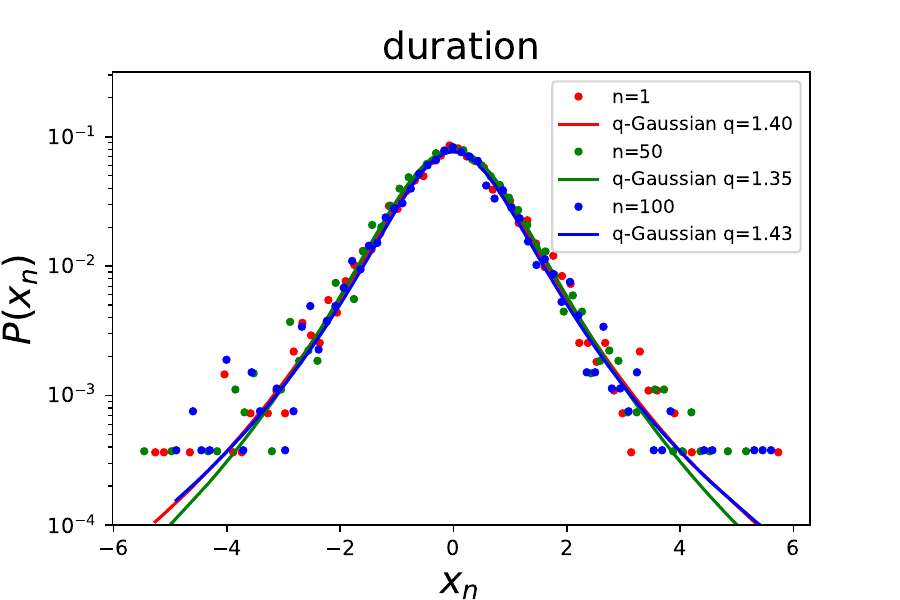}
\includegraphics[width=0.45\textwidth]{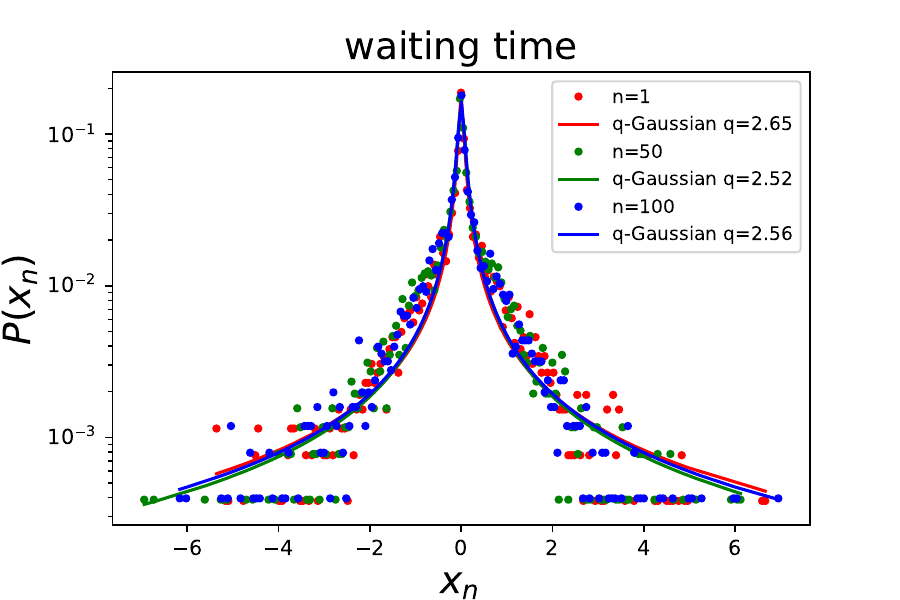}
%\vskip-0.1in
\caption{Statistical results of FRB 20201124A for the FAST sample. 
\emph{Upper-left panel}: The PDFs of energy fluctuations $X_n=S_{i+n}-S_i$ 
for different temporal interval scales $n=1$ (red dots), $n=50$ (green dots), and 
$n=100$ (blue dots). $X_n$ is rescaled by the standard deviation of $X_n$, i.e. 
$x_n=X_{n}/\sigma_{X_n}$. The data are fitted by the $q$-Gaussian function (solid curves). 
\emph{The other three panels}: The same data process and the fitting to the $q$-Gaussian function,
except now for the PDFs of fluctuations of peak flux (upper-right panel), duration (lower-left panel), 
and waiting time (lower-right panel).}
\label{fig1}
\vskip-0.2in
\end{center}
\end{figure*}

\subsection{Statistical properties of the FAST data}
Recently, \cite{Xu2022} and \cite{2022RAA....22l4002Z} reported in total 2744 bursts from the repeating source FRB 20201124A
detected by FAST from April 1 to June 11 and September 25 to October 17 in 2021. These observations were
performed in the frequency range 1.0--1.5 GHz by using the center beam of the 19-beam receiver. \cite{Xu2022} detected 
1863 bursts in 82 hours over 54 days, and \cite{2022RAA....22l4002Z} detected 881 bursts in 4 hours over 4 days. All these data 
were recorded with a frequency resolution of 122.07 kHz (i.e., 
using 4096 frequency channels to cover 1.0--1.5 GHz) and a time resolution of 49.152 $\mu$s \citep{Xu2022,2022RAA....22l4002Z,2022RAA....22l4001Z}. The total of 2744 bursts from FRB 20201124A is at present 
the largest sample observed with a single instrument and uniform selection effects. 
In this work, we use this sample to study the PDFs
of fluctuations of energy, peak flux, duration, and waiting time. FRB 20201124A was localized in a massive star-forming galaxy
with a spectroscopic redshift of $z=0.0979$ \citep{2021ApJ...919L..23F,2021A&A...656L..15P,2022MNRAS.513..982R,Xu2022}.
The known redshift allows us to infer the isotropic energy of each burst through $E=4\pi D_{L}^{2}F\Delta\nu/(1+z)$,
where $D_{L}$ is the luminosity distance, $F$ is the burst fluence, and $\Delta\nu$ is the bandwidth. The waiting time is
evaluated by the difference of times-of-arrival of two adjacent bursts.

The fluctuation of a size (e.g., energy, peak flux, duration, or waiting time) is defined as
\begin{equation}
    \label{eq:fluc}
    X_n=S_{i+n}-S_i\;,
\end{equation}
where $S_i$ is the size of the $i$-th burst (or of the $i$-th pulsar glitch, more on this below) in temporal order,
and $n$ is an arbitrary integer denoting the temporal interval scale. In order to simplify the calculation,
$X_n$ is rescaled by
\begin{equation}
    \label{eq:normal}
    x_n=\frac{X_n}{\sigma_{X_n}}\;,
\end{equation}
where $\sigma_{X_n}$ is the standard deviation of $X_n$. We will analyze the statistical properties of the dimensionless
fluctuations $x_n$.

Figure~\ref{fig1} shows the statistical results of FRB 20201124A for the FAST sample. We plot as colour dots the PDFs of
fluctuations of energy (upper-left panel), peak flux (upper-right panel), duration (lower-left panel), and waiting time
(lower-right panel) for the temporal interval scale $n=1$ (red dots), $n=50$ (green dots), and $n=100$ (blue dots). 
Empirically, the data are binned with the Freedman-Diaconis rule \citep{Freedman1981}. One can see from Figure~\ref{fig1}
that compared to a Gaussian distribution, these PDFs $P(x_{n})$ have a sharper peak at $x_{n}=0$ and fatter tails.
The sharp peak indicates that small fluctuations are most likely to occur, while fat tails indicate that there are
large but rare fluctuations. Another remarkable feature is that the data points in Figure~\ref{fig1} are almost independent
of the interval $n$ considered for the fluctuation, suggesting a common behavior of $P(x_{n})$. Here we use the Tsallis
$q$-Gaussian function \citep{1988JSP....52..479T,1998PhyA..261..534T}
\begin{equation}
    \label{eq:qgaus}
    f(x_{n})=\alpha\left[1-\beta\left(1-q\right) x_{n}^2\right]^{\frac{1}{1-q}}
\end{equation}
to fit $P(x_{n})$, where $\alpha$ is a normalization factor, the parameters $q$ and $\beta$ determine the sharpness and
width of the distribution, respectively. The $q$-Gaussian function is a generalization of the standard Gaussian function,
and it reduces to a Gaussian shape with zero mean and standard deviation $\sigma=1/\sqrt{2\beta}$ when $q$ gets close to 1.
Thus, $q\neq1$ means a departure from Gaussian behavior.

\begin{table}
\centering \caption{Fit results and estimated 68\% conﬁdence level constraints on the parameter $q$ of
the $q$-Gaussian function (Equation~(\ref{eq:qgaus})) for different temporal interval scales $n$ for repeating bursts
of FRB 20201124A (including the samples of FAST and high-energy bursts) and glitches of PSR B1737--30}
\begin{tabular}{cccc}
\hline
\hline
\multicolumn{4}{c}{FRB 20201124A} \\
\multicolumn{4}{c}{for the FAST sample} \\
   parameters  &  $n=1$   &  $n=50$  &  $n=100$     \\ 
\hline
   $q$-energy   &   $2.44^{+0.03}_{-0.03}$   &  $2.43_{-0.04}^{+0.04}$  &   $2.39_{-0.03}^{+0.03}$  \\
  $\chi^{2}_{\rm red}$ & 1.68 & 1.35 & 1.47  \\ 
 $q$-peak flux & $2.39^{+0.03}_{-0.03}$ & $2.44_{-0.03}^{+0.03}$ & $2.41_{-0.03}^{+0.03}$ \\
  $\chi^{2}_{\rm red}$ &  0.94 & 0.89 & 0.89\\ 
    $q$-duration  &  $1.40_{-0.04}^{+0.04}$  &  $1.35_{-0.04}^{+0.04}$  &  $1.43_{-0.04}^{+0.04}$   \\
        $\chi_{\rm red}^{2}$   &   0.87   &  0.77  &  1.04  \\
   $q$-waiting time  &  $2.65_{-0.03}^{+0.03}$ & $2.52_{-0.04}^{+0.04}$ & $2.56^{+0.04}_{-0.03}$ \\
   $\chi_{\rm red}^{2}$ & 1.60 & 2.03 & 1.57 \\ 
   \hline
    \hline
   \multicolumn{4}{c}{FRB 20201124A} \\
   \multicolumn{4}{c}{for the high-energy burst sample} \\
    parameters    &  $n=1$   &  $n=5$  &  $n=10$     \\ 
     \hline
   $q$-energy   &   $2.29^{+0.13}_{-0.13}$   &  $2.41_{-0.14}^{+0.13}$  &   $2.51_{-0.14}^{+0.13}$  \\
   $\chi^{2}_{\rm red}$ & 0.05 & 0.13 & 0.13  \\ 
  $q$-luminosity & $2.27^{+0.13}_{-0.15}$ & $2.40_{-0.13}^{+0.13}$ & $2.40_{-0.20}^{+0.15}$ \\
  $\chi^{2}_{\rm red}$ &  0.11 & 0.05 & 0.08\\ 
    $q$-duration  &  $2.05_{-0.20}^{+0.15}$  &  $2.19_{-0.18}^{+0.14}$  &  $1.95_{-0.25}^{+0.23}$   \\
        $\chi_{\rm red}^{2}$   &   0.09   &  0.09  &  0.15  \\
   $q$-waiting time  &  $1.95_{-0.16}^{+0.16}$ & $2.07_{-0.20}^{+0.20}$ & $2.23^{+0.28}_{-0.26}$ \\
  $\chi_{\rm red}^{2}$ & 0.18 & 0.11 & 0.13 \\ 
  \hline
  \hline
  \multicolumn{4}{c}{PSR B1737--30} \\
  parameters & $n=1$ & $n=5$ & $n=10$ \\
  \hline
    $q$-glitch size   &   $2.73^{+0.09}_{-0.11}$   &  $2.73_{-0.14}^{+0.10}$  &   $2.69_{-0.15}^{+0.12}$ \\
  $\chi^{2}_{\rm red}$ & 0.40 & 0.37 & 0.51  \\ 
   $q$-waiting time  &  $2.07_{-0.33}^{+0.30}$ & $1.84_{-0.31}^{+0.40}$ & $2.03^{+0.34}_{-0.43}$ \\
  $\chi_{\rm red}^{2}$ & 0.13 & 0.09 & 0.41 \\
\hline
\end{tabular}
\label{tab:chi3}
\tablecomments{The corresponding reduced chi-square $\chi_{\rm red}^{2}$ are also listed in the table.}
\end{table}

\begin{figure*}
\begin{center}
\vskip-0.1in
\includegraphics[width=0.45\textwidth]{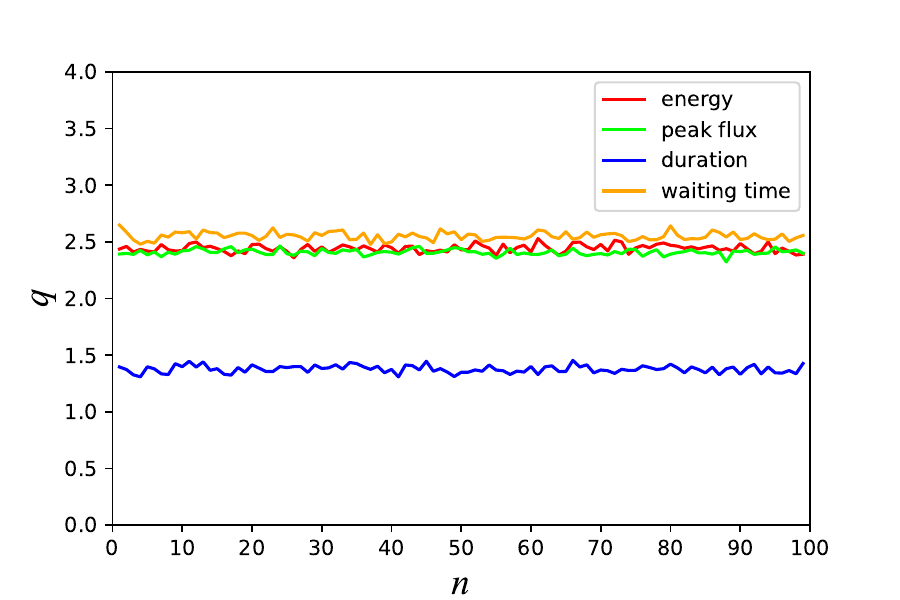}
\includegraphics[width=0.45\textwidth]{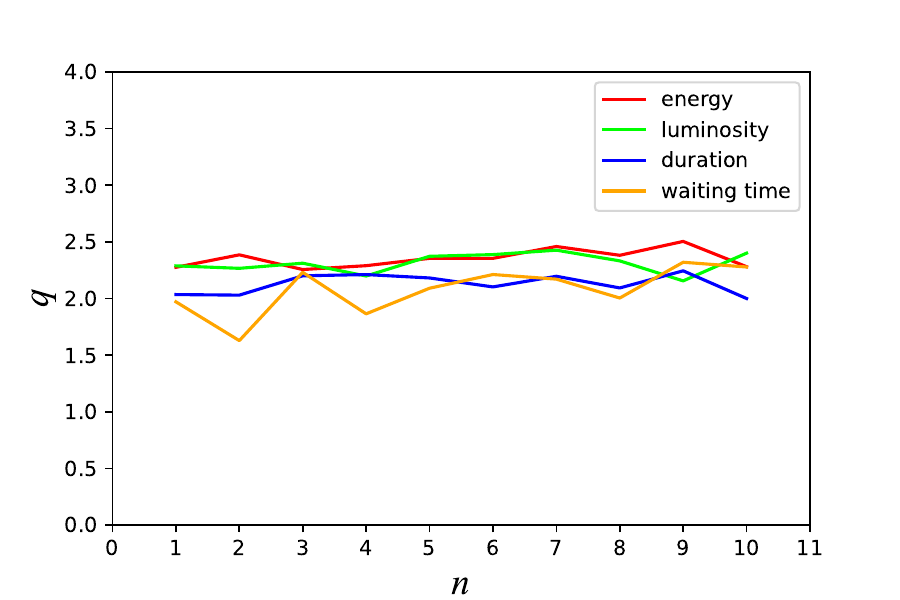}
%\vskip-0.1in
\caption{The best-fitting $q$ values in the $q$-Gaussian distribution as a function of 
the temporal interval scale $n$ for the samples of FAST (left panel) and high-energy bursts (right panel). 
The colors for different observed physical qualities (including energy, peak flux/luminosity, duration, 
and waiting time) are according to the legends.}
\label{fig2}
\vskip-0.2in
\end{center}
\end{figure*}

\begin{table*}
\centering \caption{The average $q$ values in the $q$-Gaussian distribution for FRB 20121102, FRB 20201124A, and glitches of PSR B1737--30}
\begin{tabular}{lcccccccc}
\hline
\hline
 FRB name &   Instrument  &  Energy range   &  Number  &  $q$-energy  & $q$-flux & $q$-duration & $q$-waiting time & Refs.   \\
    &      &   (erg)   &    &   &  &   &   &   \\
\hline
20121102 & FAST & $4\times10^{36}$--$8\times10^{39}$ & 1652 & $2.51\pm0.03$  & $2.50\pm0.03$ & $1.42\pm0.04$ & -- & 1\\
20201124A & FAST & $1\times10^{36}$--$3\times10^{39}$ & 2744 & $2.44\pm0.03$  & $2.40\pm0.02$ & $1.37\pm0.03$ & $2.55\pm0.03$ & 2\\
20201124A & St, O8, Tr, Wb &  $2\times10^{38}$--$2\times10^{40}$ & 46 & $2.35\pm0.07$  & $2.31\pm0.08$ & $2.13\pm0.08$ & $2.07\pm0.20$ & 2\\
\hline
PSR name &   Observatory  &  Glitch size range  &  Number  &  $q$-glitch size  &     &   & $q$-waiting time & Refs.   \\
     &      &   ($\mu$Hz)   &    &   &  &   &   &   \\
 \hline
B1737--30 &  JBO  &  $10^{-3}$--$4.4$ & 37 & $2.38\pm0.04$  &   &   & $1.80\pm0.18$ & 2\\
\hline
\end{tabular}
\label{tab2}
\tablecomments{The second column of Table~\ref{tab2} is the instrument (observatory) for observing
repeating FRBs (glitches of PSR B1737--30). The third column is the energy range (glitch size range) 
of repeating bursts of FRBs (glitches of PSR B1737--30). The fourth column represents the number of 
the corresponding sample. Columns 5--8 correspond to the average $q$ values for different observed physical qualities 
(including energy/glitch size, flux, duration, and waiting time). 
The last column is the reference list: (1) \cite{Wei2021}; (2) This work.}
\end{table*}

For a specific $n$, the free parameters ($\alpha$, $\beta$, and $q$)
can be optimized by minimizing the $\chi^2$ statistics,
\begin{equation}
    \label{eq:pdfbestchi}
    \chi^2=\sum_i \frac{\left[f\left(x_{n, i}\right)-P\left(x_{n, i}\right)\right]^2}{\sigma_{P, i}^2}\;,
\end{equation}
where $\sigma_{P,i}=\sqrt{N_{{\rm bin},i}}/N_{\rm tot}$ is the uncertainty of the data point, with $N_{{\rm bin},i}$ being
the number of $x_n$ in the $i$-th bin and $N_{\rm tot}$ being the total number of $x_n$ \citep{Wei2021}. Note that
for the sake of clarity, the error bars of the data points are not plotted in the figure. We then adopt the Python
Markov Chain Monte Carlo module, EMCEE \citep{Foreman2013}, to perform the fitting. The red, green, and blue curves
in Figure~\ref{fig1} stand for the best-fitting results for $n=1$, $n=50$, and $n=100$, respectively. 
The best-fitting $q$ values and their $1\sigma$ uncertainties for $n=1,\,50,\,100$ are listed in 
Table \ref{tab:chi3}, along with the reduced $\chi_{\rm red}^{2}$ value for the ﬁt. It is obvious that
the PDFs of fluctuations of energy, peak flux, duration, and waiting time are well reproduced by means of $q$-Gaussians.

With the FAST data, we further extract the PDFs of fluctuations of energy, peak flux, duration, and waiting time
for different scale intervals $1\le n \le100$, and fit the PDFs with the $q$-Gaussian function. The best-fitting $q$
values as a function of $n$ for energy, peak flux, duration, and waiting time are displayed in the left panel of
Figure~\ref{fig2}. One can see that the $q$ values are approximately invariant and do not depend on the scale interval $n$,
implying a scale-invariant structure of FRB 20201124A. The mean values of $q$ for energy, peak flux, duration, and
waiting time of the FAST data are $2.44\pm0.03$, $2.40\pm0.02$, $1.37\pm0.03$, and $2.55\pm0.03$, respectively,
which are listed in Table~\ref{tab2}.  Here the uncertainties represent the $1\sigma$ standard deviations of $q$ values.
Interestingly, the $q$ values we derived here are very close those of FRB 20121102 \citep{Wei2021}, which suggest that 
there is a common scale-invariant structure in repeating FRBs.

\cite{Xu2022} estimated the sample completeness for the FAST data, and showed that the fluence threshold 
achieving the 95\% detection probability with a signal-to-noise ratio $S/N>7$ is 53 mJy ms. To investigate how 
the completeness threshold would affect the modelling by $q$-Gaussians, we also perform a parallel comparative 
analysis of the complete subsample (i.e, those bursts with $F>53$ mJy ms). We find that the mean values of $q$ 
for energy, peak flux, duration, and waiting time are $2.43\pm0.03$, $2.41\pm0.02$, $1.38\pm0.03$, and $2.56\pm0.02$, 
respectively. Comparing these inferred $q$ values with those obtained from the whole FAST sample (see line 2 
in Table~\ref{tab2}), it is clear that the completeness threshold in the energy distribution only has a minimal 
influence on our results.

\subsection{Statistical properties of high-energy bursts}

Using four small 25--32-m class radio telescopes (including the 25-m telescope in Stockert, Germany (St); the 25-m
dish in Onsala, Sweden (O8); the 32-m dish in Toru\'n, Poland (Tr); and the 25-m dish in Westerbork, The Netherlands (Wb)),
\cite{Kirsten2023} detected 46 high-energy bursts from FRB 20201124A. They concluded that the highest-energy bursts
($E>3\times 10^{39}$ erg) occur much more frequently than one would expect based on the energy distribution of
lower-energy bursts observed by FAST. The hyperactive source FRB 20201124A provides a good opportunity to probe
the high-energy distribution and to compare with low-energy bursts. Here we investigate whether the highest-energy
bursts originate from a separate emission mechanism by comparing the scale-invariant properties of low- and high-energy bursts.

Due to the limited number of data points, the cumulative distribution function (CDF) is often used instead of PDF to avoid
the arbitrariness caused by binning. We thus try to fit the CDF of fluctuations for high-energy bursts using the CDF of
$q$-Gaussian, i.e.,
\begin{equation}
    \label{eq:cdf}
    F(x_{n})=\int_{-\infty}^{x_{n}} f\left(x_{n}\right) {\rm d} x\;,
\end{equation}
where $f(x_{n})$ is the $q$-Gaussian function (Equation~(\ref{eq:qgaus})).
Similarly, with a fixed $n$, we obtain the best-fitting parameters ($\alpha$, $\beta$, and $q$) by minimizing the $\chi^2$ statistics,
\begin{equation}
    \label{eq:cdfchi2}
    \chi^2=\sum_i \frac{\left[N_{\mathrm{cum}}\left(<x_{n, i}\right)-F\left(x_{n, i}\right)\right]^2}{\sigma_{\text{cum}, i}^2}\;,
\end{equation}
where $\sigma _{{\rm cum},i}=\sqrt{N_{\rm cum}(<x_{n,i})}$ denotes the uncertainty of the data point, with $N_{\rm cum}(<x_{n,i})$
being the cumulative number of the fluctuations \citep{Wei2023}.

\begin{figure*}
\begin{center}
\vskip-0.1in
\includegraphics[width=0.45\textwidth]{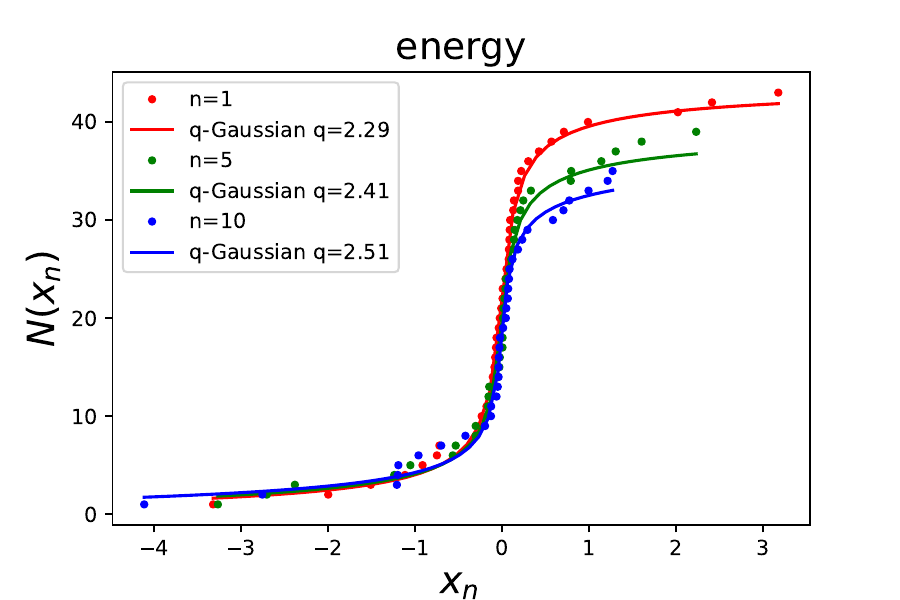}
\includegraphics[width=0.45\textwidth]{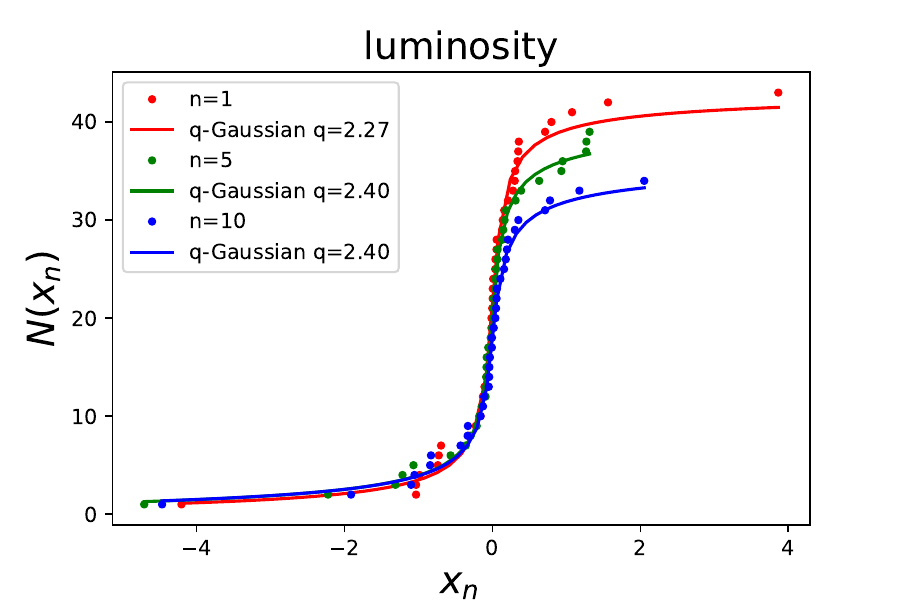}
\includegraphics[width=0.45\textwidth]{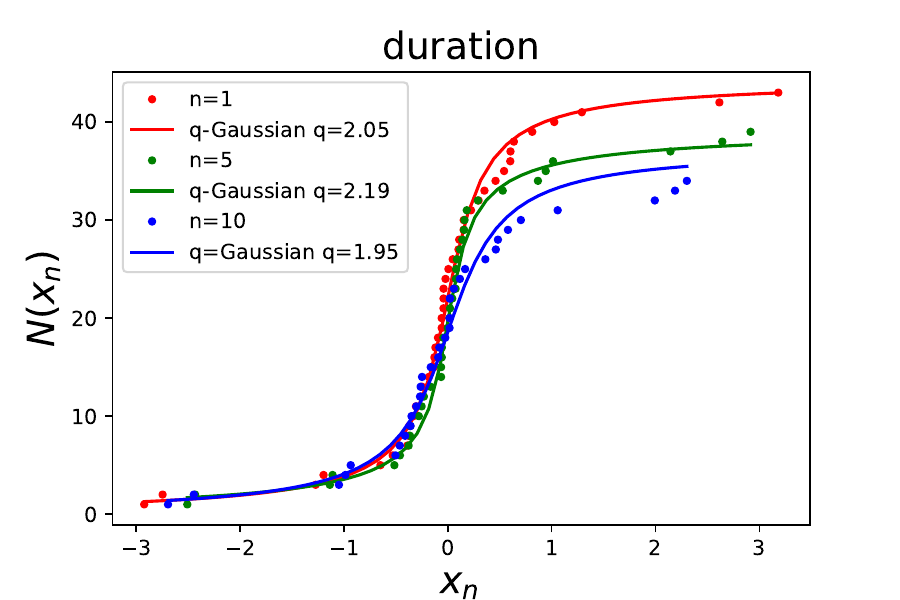}
\includegraphics[width=0.45\textwidth]{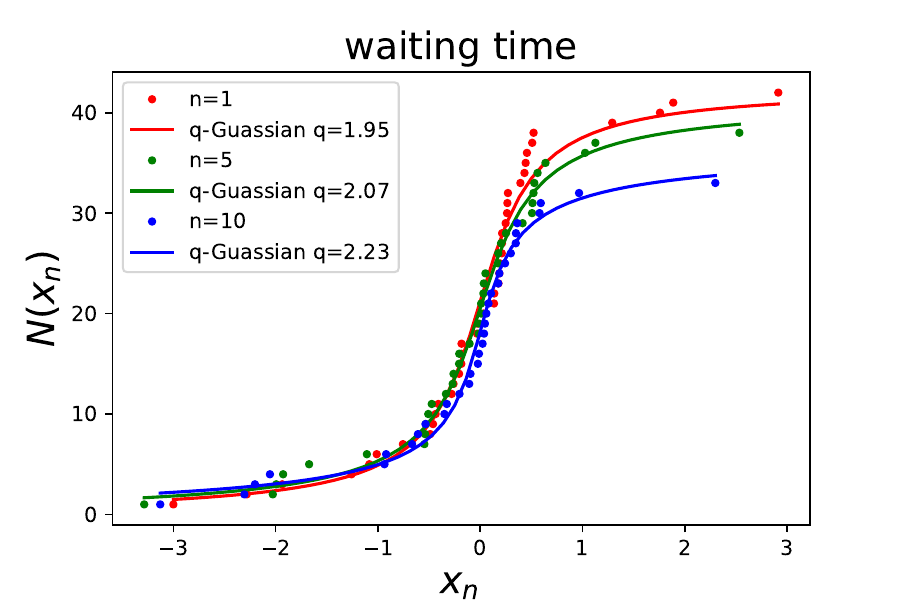}
%\vskip-0.1in
\caption{Statistical results of FRB 20201124A for the high-energy burst sample. 
\emph{Upper-left panel}: The CDFs of energy fluctuations $X_n=S_{i+n}-S_i$ 
for different temporal interval scales $n=1$ (red dots), $n=5$ (green dots), and 
$n=10$ (blue dots). $X_n$ is rescaled by the standard deviation of $X_n$, i.e. 
$x_n=X_{n}/\sigma_{X_n}$. The data are fitted by the CDF of $q$-Gaussian function (solid curves). 
\emph{The other three panels}: The same data process and the fitting to the CDF of $q$-Gaussian function,
except now for the CDFs of fluctuations of luminosity (upper-right panel), duration (lower-left panel), 
and waiting time (lower-right panel).}
\label{fig3}
\vskip-0.2in
\end{center}
\end{figure*}

Figure~\ref{fig3} shows some examples of the fits for the high-energy burst sample. In this plot, we display the CDFs
of fluctuations of energy (upper-left panel), luminosity (upper-right panel), duration (lower-left panel), and waiting time
(lower-right panel) for $n=1$ (red dots), $n=5$ (green dots), and $n=10$ (blue dots). The best fitting 
$q$-values and the corresponding $\chi^2_{\rm red}$ values for $n=1,\,5,\,10$ are presented in Table \ref{tab:chi3}.
From Figure~\ref{fig3} and the $\chi^2_{\rm red}$ values, we find that the CDFs of fluctuations of energy, luminosity, 
duration, and waiting time can be well fitted by the CDF of $q$-Gaussian function (see solid curves). 
Moreover, we plot the best-fitting $q$ values as a function of the temporal interval scale $n$ in the range $1\le n \le 10$,
and find that the $q$ values also keep approximately steady (see the right panel of Figure~\ref{fig2}). The average $q$ values
of energy, luminosity, duration, and waiting time for the data of high-energy bursts are $2.35\pm0.07$, $2.31\pm0.08$,
$2.13\pm0.08$, and $2.07\pm0.20$, respectively. As shown in Table~\ref{tab2}, the $q$ values of energy and luminosity
(peak flux) of high-energy bursts are consistent with those of FAST observations of low-energy bursts within 1$\sigma$
confidence level, but the $q$ values of duration and waiting time between low- and high-energy bursts are significantly
different. 

One possible explanation for the different $q$ values of duration and waiting time for low- and high-energy
bursts of FRB 20201124A is that burst duration is subject to instrumental effects, thereby affecting the estimate of 
the waiting time. 
It is well known that the observed burst duration would be broadened by instrumental effects \citep{2003ApJ...596.1142C}. 
The instrumental burst broadening includes the intrachannel dispersion smearing and the sampling timescale. The smearing 
is due to intrachannel dispersion. The time resolution should not be better than the sampling timescale, which broadens 
the burst. Therefore, different instruments may correspond to different broadening components. However, since we are 
interested in the statistical distributions of the differences of durations and waiting times at different time intervals, 
burst broadening induced by instrumental effects can be approximately deducted from subtracting two observed durations 
(or two waiting times). Thus, we conclude that the different $q$ values for low- and high-energy bursts is likely not 
an instrumental effect. Rather, it more likely hints a differing emission mechanism or emission site between low- and 
high-energy bursts, although they originate from the same progenitor source.

\section{Scale Invariance in Glitching Pulsars}
\label{sec:glitch}

\begin{figure*}
\begin{center}
\vskip-0.1in
\includegraphics[width=0.45\textwidth]{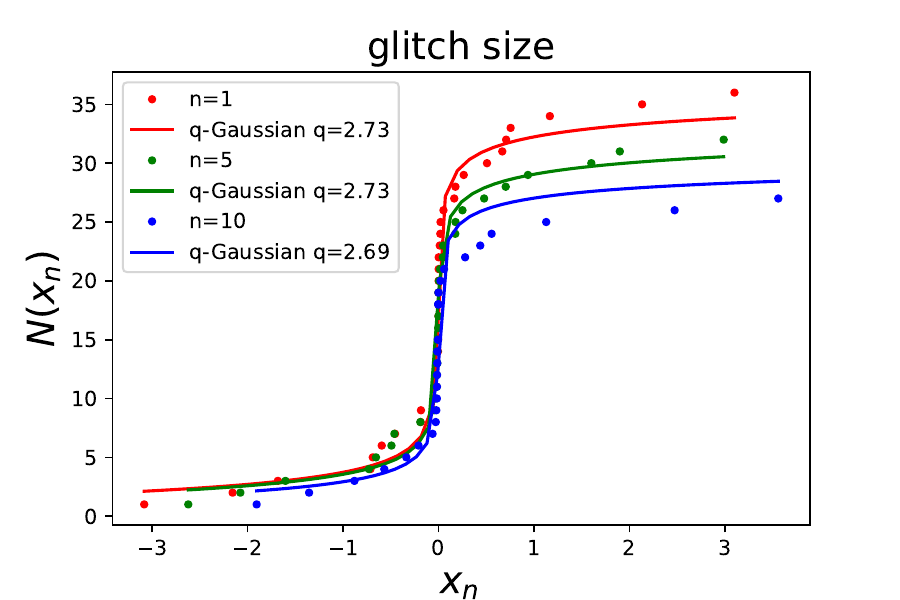}
\includegraphics[width=0.45\textwidth]{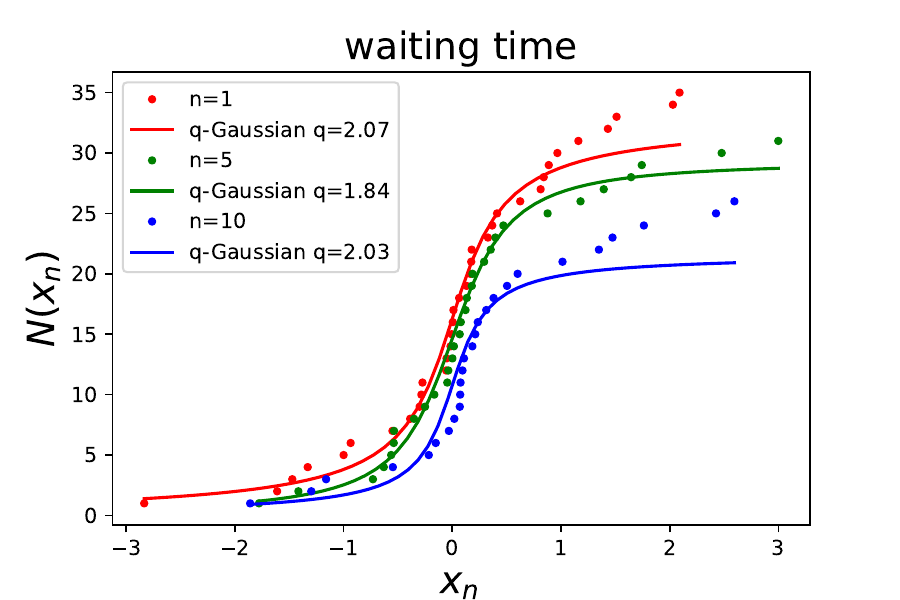}
\includegraphics[width=0.45\textwidth]{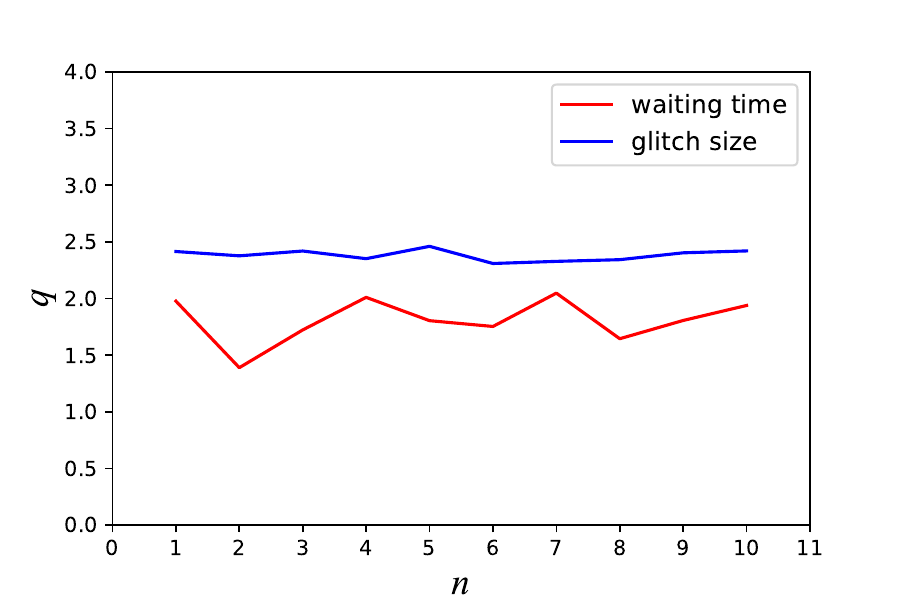}
%\vskip-0.1in
\caption{Statistical results of glitches of PSR B1737--30. 
\emph{Upper-left panel}: The CDFs of glitch size fluctuations $X_n=S_{i+n}-S_i$ 
for different temporal interval scales $n=1$ (red dots), $n=5$ (green dots), and 
$n=10$ (blue dots). $X_n$ is rescaled by the standard deviation of $X_n$, i.e. 
$x_n=X_{n}/\sigma_{X_n}$. The data are fitted by the CDF of $q$-Gaussian function (solid curves). 
\emph{Upper-right panel}: The same data process and the fitting to the CDF of $q$-Gaussian function,
except now for the CDFs of waiting-time fluctuations. \emph{Lower panel}}: The best-fitting 
$q$ values in the $q$-Gaussian distribution as a function of $n$.
\label{fig4}
\vskip-0.2in
\end{center}
\end{figure*}

Remarkably, both a glitch and an antiglitch accompanied by FRB-like bursts from SGR J1935+2154 were
recently reported. The temporal coincidences between the glitch/antiglitch and FRB-like bursts imply some physical
connection between them \citep{2022arXiv221103246G,2023NatAs...7..339Y}. The statistical distributions of glitch sizes
($\Delta \nu$) and times between successive glitches (waiting times) for some individual pulsars are power-law forms
\citep{1993EL.....23..185M,2008ApJ...672.1103M,2021ApJ...917....1C}. Power-law distributions of energy and waiting time
are also found in some repeating FRBs \citep{2017JCAP...03..023W,2018MNRAS.476.1849K,2019gour,2020MNRAS.491.1498C,
2021cruces,2021Natur.598..267L,2021ApJ...920L..23Z,2022MNRAS.515.3577H,2023MNRAS.519..666J,2023ApJ...949L..33W}. 
Inspired by the associations between the glitch/antiglitch
and FRB-like bursts of SGR J1935+2154 and the statistical similarities between pulsar glitches and repeating FRBs,
here we investigate whether pulsar glitches have a similar scale-invariant behavior.

To date, there are only eight pulsars with more than 10 detected glitches \citep{2019A&A...630A.115F}. PSR B1737--30
has been observed regularly by the Jodrell Bank Observatory (JBO), and 37 glitches have been detected over $\sim 35$ years.
PSR B1737--30 has the largest number of glitch events among the glitching pulsars exhibiting power-law size distributions.
Consequently, the statistical results for PSR B1737--30 are much more significant than other pulsars.
Glitch epochs and sizes for the pulsars are available in the JBO online catalogue
\citep{Espinoza2011}\footnote{\url{http://www.jb.man.ac.uk/pulsar/glitches.html}}. Due to the relatively small sample
of glitches, we also focus on the CDFs of fluctuations of glitch size and waiting time at different scale intervals $n$.
For a fixed $n$, we use the CDF of $q$-Gaussian (Equation~(\ref{eq:cdf})) to fit the CDFs of fluctuations, and derive
the best-fitting $q$ value. Then we vary $n$ and obtain the $q$ values as a function of $n$. 

For the detected glitches in PSR B1737--30, the CDFs of fluctuations of glitch size and waiting time are presented
in the left and middle panels of Figure~\ref{fig4}, respectively. The best-fitting $q$ values and 
the corresponding $\chi_{\rm red}^2$ values for $n=1,\,5,\,10$ are listed in Table \ref{tab:chi3}.
It is clear that the $q$-Gaussian function fits
the data points very well. To better highlight the scale-invariant property in PSR B1737--30' glitches,
the best-fitting $q$ values as a function of $n$ for glitch size and waiting time are plotted in the right panel
of Figure~\ref{fig4}. Again, we find that the $q$ values are nearly constant and independent of $n$. We also list
the average $q$ values from $n=1$ to 10 for glitch size ($\overline{q}=2.38\pm0.04$) and waiting time
($\overline{q}=1.80\pm0.18$) in Table~\ref{tab2}. Interestingly, the average $q$ values of glitch size (energy)
and waiting time are well consistent with those of FRB 20201124A's high-energy bursts, which indicate that
there may be an underlying physical connection between pulsar glitches and FRB 20201124A's high-energy bursts.

\section{Summary and discussion}
\label{sec:sum}

SOC dynamics can be effectively identified and diagnosed by analyzing the avalanche size fluctuations \citep{Caruso2007}.
When criticality appears, the PDFs for the avalanche size fluctuations at different times have fat tails with a $q$-Gaussian
form. Such a $q$-Gaussian behavior is independent of the time interval adopted, and it is found so when considering energy
fluctuations between real earthquakes \citep{Caruso2007}. That is, the $q$ values in $q$-Gaussian distributions are nearly
invariant for different temporal interval scales, implying a scale-invariant structure in earthquakes 
(see also \citealt{Wang2015}).
Analogous scale-invariant characteristics have also been discovered in some astronomical phenomena, such as SGRs
\citep{Chang2017,Wei2021,2022MNRAS.510.1801S}, the repeating FRB 20121102 \citep{Lin2020,Wei2021}, GRB X-ray flares
\citep{Wei2023} and precursors \citep{2023ApJ...955L..34L}, and gamma-ray flares of the Sun and the blazar 3C 454.3
\citep{2023ApJ...959..109P}. Given the temporal coincidences between the glitch/antiglitch
and FRB-like bursts from the Galactic magnetar SGR J1935+2154 \citep{2022arXiv221103246G,2023NatAs...7..339Y}, here
we examined their possible physical connection by comparing the scale-invariant similarities between repeating FRBs
and glitching pulsars.

For repeating FRBs, we focused on the statistical properties of the hyperactive repeating source FRB 20201124A
using two samples from different observations. The first sample contains 2744 bursts from the FAST observation,
and the second one consists of 46 high-energy bursts from the observations of four small 25--32-m class radio telescopes.
Note that except for FRB 20121102, it is unclear whether other repeating FRBs share the same scale-invariant property.
With the FAST data, we confirmed that the PDFs of fluctuations of energy, peak flux, duration, and waiting time can be
well fitted by a $q$-Gaussian function, and the $q$ values keep nearly steady for different time scales. Moreover,
we found that the average $q$ values of energy, peak flux, and duration for the FAST data of FRB 20201124A are
very close to those of FRB 20121102 \citep{Wei2021}, which indicate that there is a common scale-invariant feature
in repeating FRBs. With the high-energy burst sample, similar scale-invariant results were obtained, but the average
$q$ values of duration and waiting time of high-energy bursts are significantly different from those of FAST
observations of low-energy bursts. This implies that low- and high-energy bursts may originate from different
emission mechanisms or emission regions at the progenitor source. \cite{Wang2015} also found that different
faulting styles correspond to different $q$ values in earthquakes.

For glitching pulsars, we investigated the statistical properties of 37 known glitches from PSR B1737--30.
We showed that the distributions of fluctuations of glitch size and waiting time also exhibit a $q$-Gaussian
form, with constant $q$ values independent of the adopted time scale. This scale-invariant property is very
similar to those of repeating FRBs, and both the two astronomical phenomena can be attributed to a SOC process.
Interestingly, the average $q$ values of glitch size (energy) and waiting time align consistently with those of
FRB 20201124A's high-energy bursts, which suggest that there may be an underlying physical association between
pulsar glitches and FRB 20201124A's high-energy bursts.

In summary, our findings support the argument that both repeating FRBs and pulsar glitches can be explained
within a dissipative SOC mechanism with long-range interactions. In the future, much more glitches and FRB-like bursts
from the magnetars will be detected. The physical connection between glitches and FRB-like bursts can be further
investigated.

\begin{acknowledgments}
We are grateful to the anonymous referee for constructive comments that helped improve our work.
This work is partially supported by the National SKA Program of
China (2022SKA0130100), the National Natural Science Foundation of China (grant Nos. 12373053, 12321003, and
12041306), the Key Research Program of Frontier Sciences (grant No. ZDBS-LY-7014)
of Chinese Academy of Sciences, International Partnership Program of Chinese Academy of Sciences
for Grand Challenges (114332KYSB20210018), the CAS Project for Young Scientists in Basic Research
(grant No. YSBR-063), the CAS Organizational Scientific Research Platform for National Major
Scientific and Technological Infrastructure: Cosmic Transients with FAST, and the Natural Science
Foundation of Jiangsu Province (grant No. BK20221562).
\end{acknowledgments}

\bibliography{sample631}{}

\begin{thebibliography}{}
\expandafter\ifx\csname natexlab\endcsname\relax\def\natexlab#1{#1}\fi
\providecommand{\url}[1]{\href{#1}{#1}}
\providecommand{\dodoi}[1]{doi:~\href{http://doi.org/#1}{\nolinkurl{#1}}}
\providecommand{\doeprint}[1]{\href{http://ascl.net/#1}{\nolinkurl{http://ascl.net/#1}}}
\providecommand{\doarXiv}[1]{\href{https://arxiv.org/abs/#1}{\nolinkurl{https://arxiv.org/abs/#1}}}

\bibitem[{{Antonopoulou} {et~al.}(2022){Antonopoulou}, {Haskell}, \& {Espinoza}}]{2022RPPh...85l6901A}
{Antonopoulou}, D., {Haskell}, B., \& {Espinoza}, C.~M. 2022, Reports on Progress in Physics, 85, 126901, \dodoi{10.1088/1361-6633/ac9ced}

\bibitem[{{Aschwanden}(2011)}]{2011soca.book.....A}
{Aschwanden}, M.~J. 2011, {Self-Organized Criticality in Astrophysics}

\bibitem[{{Aschwanden}(2012)}]{2012A&A...539A...2A}
---. 2012, \aap, 539, A2, \dodoi{10.1051/0004-6361/201118237}

\bibitem[{{Aschwanden}(2014)}]{2014ApJ...782...54A}
---. 2014, \apj, 782, 54, \dodoi{10.1088/0004-637X/782/1/54}

\bibitem[{{Aschwanden}(2015)}]{2015ApJ...814...19A}
---. 2015, \apj, 814, 19, \dodoi{10.1088/0004-637X/814/1/19}

\bibitem[{{Aschwanden} {et~al.}(2016){Aschwanden}, {Crosby}, {Dimitropoulou}, {Georgoulis}, {Hergarten}, {McAteer}, {Milovanov}, {Mineshige}, {Morales}, {Nishizuka}, {Pruessner}, {Sanchez}, {Sharma}, {Strugarek}, \& {Uritsky}}]{2016SSRv..198...47A}
{Aschwanden}, M.~J., {Crosby}, N.~B., {Dimitropoulou}, M., {et~al.} 2016, \ssr, 198, 47, \dodoi{10.1007/s11214-014-0054-6}

\bibitem[{Bak {et~al.}(1987)Bak, Tang, \& Wiesenfeld}]{bak1987}
Bak, P., Tang, C., \& Wiesenfeld, K. 1987, Physical Review Letters, 59, 381, \dodoi{10.1103/PhysRevLett.59.381}

\bibitem[{{Bochenek} {et~al.}(2020){Bochenek}, {Ravi}, {Belov}, {Hallinan}, {Kocz}, {Kulkarni}, \& {McKenna}}]{2020Natur.587...59B}
{Bochenek}, C.~D., {Ravi}, V., {Belov}, K.~V., {et~al.} 2020, \nat, 587, 59, \dodoi{10.1038/s41586-020-2872-x}

\bibitem[{{Carlin} \& {Melatos}(2021)}]{2021ApJ...917....1C}
{Carlin}, J.~B., \& {Melatos}, A. 2021, \apj, 917, 1, \dodoi{10.3847/1538-4357/ac06a2}

\bibitem[{{Caruso} {et~al.}(2007){Caruso}, {Pluchino}, {Latora}, {Vinciguerra}, \& {Rapisarda}}]{Caruso2007}
{Caruso}, F., {Pluchino}, A., {Latora}, V., {Vinciguerra}, S., \& {Rapisarda}, A. 2007, \pre, 75, 055101, \dodoi{10.1103/PhysRevE.75.055101}

\bibitem[{{Chang} {et~al.}(2017){Chang}, {Lin}, {Sang}, \& {Wang}}]{Chang2017}
{Chang}, Z., {Lin}, H.-N., {Sang}, Y., \& {Wang}, P. 2017, Chinese Physics C, 41, 065104, \dodoi{10.1088/1674-1137/41/6/065104}

\bibitem[{{Cheng} {et~al.}(2020){Cheng}, {Zhang}, \& {Wang}}]{2020MNRAS.491.1498C}
{Cheng}, Y., {Zhang}, G.~Q., \& {Wang}, F.~Y. 2020, \mnras, 491, 1498, \dodoi{10.1093/mnras/stz3085}

\bibitem[{{CHIME/FRB Collaboration} {et~al.}(2020{\natexlab{a}}){CHIME/FRB Collaboration}, {Amiri}, {Andersen}, {Bandura}, {Bhardwaj}, {Boyle}, {Brar}, {Chawla}, {Chen}, {Cliche}, {Cubranic}, {Deng}, {Denman}, {Dobbs}, {Dong}, {Fandino}, {Fonseca}, {Gaensler}, {Giri}, {Good}, {Halpern}, {Hessels}, {Hill}, {H{\"o}fer}, {Josephy}, {Kania}, {Karuppusamy}, {Kaspi}, {Keimpema}, {Kirsten}, {Landecker}, {Lang}, {Leung}, {Li}, {Lin}, {Marcote}, {Masui}, {McKinven}, {Mena-Parra}, {Merryfield}, {Michilli}, {Milutinovic}, {Mirhosseini}, {Naidu}, {Newburgh}, {Ng}, {Nimmo}, {Paragi}, {Patel}, {Pen}, {Pinsonneault-Marotte}, {Pleunis}, {Rafiei-Ravandi}, {Rahman}, {Ransom}, {Renard}, {Sanghavi}, {Scholz}, {Shaw}, {Shin}, {Siegel}, {Singh}, {Smegal}, {Smith}, {Stairs}, {Tendulkar}, {Tretyakov}, {Vanderlinde}, {Wang}, {Wang}, {Wulf}, {Yadav}, \& {Zwaniga}}]{chime2020}
{CHIME/FRB Collaboration}, {Amiri}, M., {Andersen}, B.~C., {et~al.} 2020{\natexlab{a}}, \nat, 582, 351, \dodoi{10.1038/s41586-020-2398-2}

\bibitem[{{CHIME/FRB Collaboration} {et~al.}(2020{\natexlab{b}}){CHIME/FRB Collaboration}, {Andersen}, {Bandura}, {Bhardwaj}, {Bij}, {Boyce}, {Boyle}, {Brar}, {Cassanelli}, {Chawla}, {Chen}, {Cliche}, {Cook}, {Cubranic}, {Curtin}, {Denman}, {Dobbs}, {Dong}, {Fandino}, {Fonseca}, {Gaensler}, {Giri}, {Good}, {Halpern}, {Hill}, {Hinshaw}, {H{\"o}fer}, {Josephy}, {Kania}, {Kaspi}, {Landecker}, {Leung}, {Li}, {Lin}, {Masui}, {McKinven}, {Mena-Parra}, {Merryfield}, {Meyers}, {Michilli}, {Milutinovic}, {Mirhosseini}, {M{\"u}nchmeyer}, {Naidu}, {Newburgh}, {Ng}, {Patel}, {Pen}, {Pinsonneault-Marotte}, {Pleunis}, {Quine}, {Rafiei-Ravandi}, {Rahman}, {Ransom}, {Renard}, {Sanghavi}, {Scholz}, {Shaw}, {Shin}, {Siegel}, {Singh}, {Smegal}, {Smith}, {Stairs}, {Tan}, {Tendulkar}, {Tretyakov}, {Vanderlinde}, {Wang}, {Wulf}, \& {Zwaniga}}]{2020Natur.587...54C}
{CHIME/FRB Collaboration}, {Andersen}, B.~C., {Bandura}, K.~M., {et~al.} 2020{\natexlab{b}}, \nat, 587, 54, \dodoi{10.1038/s41586-020-2863-y}

\bibitem[{{Cordes} \& {McLaughlin}(2003)}]{2003ApJ...596.1142C}
{Cordes}, J.~M., \& {McLaughlin}, M.~A. 2003, \apj, 596, 1142, \dodoi{10.1086/378231}

\bibitem[{{Cruces} {et~al.}(2021){Cruces}, {Spitler}, {Scholz}, {Lynch}, {Seymour}, {Hessels}, {Gouiff{\'e}s}, {Hilmarsson}, {Kramer}, \& {Munjal}}]{2021cruces}
{Cruces}, M., {Spitler}, L.~G., {Scholz}, P., {et~al.} 2021, \mnras, 500, 448, \dodoi{10.1093/mnras/staa3223}

\bibitem[{Espinoza {et~al.}(2011)Espinoza, Lyne, Stappers, \& Kramer}]{Espinoza2011}
Espinoza, C.~M., Lyne, A.~G., Stappers, B.~W., \& Kramer, M. 2011, Monthly Notices of the Royal Astronomical Society, 414, 1679, \dodoi{10.1111/j.1365-2966.2011.18503.x}

\bibitem[{{Fong} {et~al.}(2021){Fong}, {Dong}, {Leja}, {Bhandari}, {Day}, {Deller}, {Kumar}, {Prochaska}, {Scott}, {Bannister}, {Eftekhari}, {Gordon}, {Heintz}, {James}, {Kilpatrick}, {Mahony}, {Rouco Escorial}, {Ryder}, {Shannon}, \& {Tejos}}]{2021ApJ...919L..23F}
{Fong}, W.-f., {Dong}, Y., {Leja}, J., {et~al.} 2021, \apjl, 919, L23, \dodoi{10.3847/2041-8213/ac242b}

\bibitem[{{Foreman-Mackey} {et~al.}(2013){Foreman-Mackey}, {Hogg}, {Lang}, \& {Goodman}}]{Foreman2013}
{Foreman-Mackey}, D., {Hogg}, D.~W., {Lang}, D., \& {Goodman}, J. 2013, \pasp, 125, 306, \dodoi{10.1086/670067}

\bibitem[{{Freedman} \& {Diaconis}(1981)}]{Freedman1981}
{Freedman}, D., \& {Diaconis}, P. 1981, Probability Theory and Related Fields, 57, 453, \dodoi{https://doi.org/10.1007/BF01025868}

\bibitem[{{Fuentes} {et~al.}(2019){Fuentes}, {Espinoza}, \& {Reisenegger}}]{2019A&A...630A.115F}
{Fuentes}, J.~R., {Espinoza}, C.~M., \& {Reisenegger}, A. 2019, \aap, 630, A115, \dodoi{10.1051/0004-6361/201935939}

\bibitem[{{Ge} {et~al.}(2022){Ge}, {Yang}, {Lu}, {Zhou}, {Ji}, {Zhang}, {Zhang}, {Zhang}, {Wang}, {Lee}, {Zhu}, {Li}, {Hou}, \& {Li}}]{2022arXiv221103246G}
{Ge}, M., {Yang}, Y.-P., {Lu}, F., {et~al.} 2022, arXiv e-prints, arXiv:2211.03246, \dodoi{10.48550/arXiv.2211.03246}

\bibitem[{{Good} \& {Chime/Frb Collaboration}(2020)}]{2020ATel14074....1G}
{Good}, D., \& {Chime/Frb Collaboration}. 2020, The Astronomer's Telegram, 14074, 1

\bibitem[{{Gourdji} {et~al.}(2019){Gourdji}, {Michilli}, {Spitler}, {Hessels}, {Seymour}, {Cordes}, \& {Chatterjee}}]{2019gour}
{Gourdji}, K., {Michilli}, D., {Spitler}, L.~G., {et~al.} 2019, \apjl, 877, L19, \dodoi{10.3847/2041-8213/ab1f8a}

\bibitem[{{Haskell} \& {Melatos}(2015)}]{2015IJMPD..2430008H}
{Haskell}, B., \& {Melatos}, A. 2015, International Journal of Modern Physics D, 24, 1530008, \dodoi{10.1142/S0218271815300086}

\bibitem[{{Hewitt} {et~al.}(2022){Hewitt}, {Snelders}, {Hessels}, {Nimmo}, {Jahns}, {Spitler}, {Gourdji}, {Hilmarsson}, {Michilli}, {Ould-Boukattine}, {Scholz}, \& {Seymour}}]{2022MNRAS.515.3577H}
{Hewitt}, D.~M., {Snelders}, M.~P., {Hessels}, J.~W.~T., {et~al.} 2022, \mnras, 515, 3577, \dodoi{10.1093/mnras/stac1960}

\bibitem[{{Huang} \& {Geng}(2014)}]{2014ApJ...782L..20H}
{Huang}, Y.~F., \& {Geng}, J.~J. 2014, \apjl, 782, L20, \dodoi{10.1088/2041-8205/782/2/L20}

\bibitem[{{Jahns} {et~al.}(2023){Jahns}, {Spitler}, {Nimmo}, {Hewitt}, {Snelders}, {Seymour}, {Hessels}, {Gourdji}, {Michilli}, \& {Hilmarsson}}]{2023MNRAS.519..666J}
{Jahns}, J.~N., {Spitler}, L.~G., {Nimmo}, K., {et~al.} 2023, \mnras, 519, 666, \dodoi{10.1093/mnras/stac3446}

\bibitem[{Katz(1986)}]{Katz1986}
Katz, J.~I. 1986, Journal of Geophysical Research, 91, 10412, \dodoi{10.1029/jb091ib10p10412}

\bibitem[{{Katz}(2018)}]{2018MNRAS.476.1849K}
{Katz}, J.~I. 2018, \mnras, 476, 1849, \dodoi{10.1093/mnras/sty366}

\bibitem[{{Kirsten} {et~al.}(2024){Kirsten}, {Ould-Boukattine}, {Herrmann}, {Gawro{\'n}ski}, {Hessels}, {Lu}, {Snelders}, {Chawla}, {Yang}, {Blaauw}, {Nimmo}, {Puchalska}, {Wolak}, \& {van Ruiten}}]{Kirsten2023}
{Kirsten}, F., {Ould-Boukattine}, O.~S., {Herrmann}, W., {et~al.} 2024, Nature Astronomy, \dodoi{10.1038/s41550-023-02153-z}

\bibitem[{{Lanman} {et~al.}(2022){Lanman}, {Andersen}, {Chawla}, {Josephy}, {Noble}, {Kaspi}, {Bandura}, {Bhardwaj}, {Boyle}, {Brar}, {Breitman}, {Cassanelli}, {Dong}, {Fonseca}, {Gaensler}, {Good}, {Kaczmarek}, {Leung}, {Masui}, {Meyers}, {Ng}, {Patel}, {Pearlman}, {Petroff}, {Pleunis}, {Rafiei-Ravandi}, {Rahman}, {Sanghavi}, {Scholz}, {Shin}, {Stairs}, {Tendulkar}, \& {Zwaniga}}]{2022ApJ...927...59L}
{Lanman}, A.~E., {Andersen}, B.~C., {Chawla}, P., {et~al.} 2022, \apj, 927, 59, \dodoi{10.3847/1538-4357/ac4bc7}

\bibitem[{{Li} {et~al.}(2021{\natexlab{a}}){Li}, {Lin}, {Xiong}, {Ge}, {Li}, {Li}, {Lu}, {Zhang}, {Tuo}, {Nang}, {Zhang}, {Xiao}, {Chen}, {Song}, {Xu}, {Liu}, {Jia}, {Cao}, {Qu}, {Zhang}, {Gu}, {Liao}, {Zhao}, {Tan}, {Nie}, {Zhao}, {Zheng}, {Zheng}, {Luo}, {Cai}, {Li}, {Xue}, {Bu}, {Chang}, {Chen}, {Chen}, {Chen}, {Chen}, {Chen}, {Cui}, {Cui}, {Deng}, {Dong}, {Du}, {Fu}, {Gao}, {Gao}, {Gao}, {Gu}, {Guan}, {Guo}, {Han}, {Huang}, {Huo}, {Jiang}, {Jiang}, {Jin}, {Jin}, {Kong}, {Li}, {Li}, {Li}, {Li}, {Li}, {Li}, {Li}, {Liang}, {Liu}, {Liu}, {Liu}, {Liu}, {Liu}, {Lu}, {Lu}, {Luo}, {Ma}, {Meng}, {Ou}, {Sai}, {Shang}, {Song}, {Sun}, {Tao}, {Wang}, {Wang}, {Wang}, {Wang}, {Wang}, {Wen}, {Wu}, {Wu}, {Wu}, {Xiao}, {Xu}, {Yang}, {Yang}, {Yang}, {Yang}, {Yi}, {Yin}, {You}, {Zhang}, {Zhang}, {Zhang}, {Zhang}, {Zhang}, {Zhang}, {Zhang}, {Zhang}, {Zhang}, {Zhang}, {Zhang}, {Zhang}, {Zhang}, {Zhang}, {Zhang}, {Zhang}, {Zhou}, {Zhou}, {Zhu}, {Zhu}, \& {Zhuang}}]{2021NatAs...5..378L}
{Li}, C.~K., {Lin}, L., {Xiong}, S.~L., {et~al.} 2021{\natexlab{a}}, Nature Astronomy, 5, 378, \dodoi{10.1038/s41550-021-01302-6}

\bibitem[{{Li} {et~al.}(2021{\natexlab{b}}){Li}, {Wang}, {Zhu}, {Zhang}, {Zhang}, {Duan}, {Zhang}, {Feng}, {Tang}, {Chatterjee}, {Cordes}, {Cruces}, {Dai}, {Gajjar}, {Hobbs}, {Jin}, {Kramer}, {Lorimer}, {Miao}, {Niu}, {Niu}, {Pan}, {Qian}, {Spitler}, {Werthimer}, {Zhang}, {Wang}, {Xie}, {Yue}, {Zhang}, {Zhi}, \& {Zhu}}]{2021Natur.598..267L}
{Li}, D., {Wang}, P., {Zhu}, W.~W., {et~al.} 2021{\natexlab{b}}, \nat, 598, 267, \dodoi{10.1038/s41586-021-03878-5}

\bibitem[{{Li} \& {Yang}(2023)}]{2023ApJ...955L..34L}
{Li}, X.-J., \& {Yang}, Y.-P. 2023, \apjl, 955, L34, \dodoi{10.3847/2041-8213/acf12c}

\bibitem[{Lin \& Sang(2020)}]{Lin2020}
Lin, H.~N., \& Sang, Y. 2020, Monthly Notices of the Royal Astronomical Society, 491, 2156, \dodoi{10.1093/mnras/stz3149}

\bibitem[{Lorimer {et~al.}(2007)Lorimer, Bailes, McLaughlin, Narkevic, \& Crawford}]{lorimer2007}
Lorimer, D.~R., Bailes, M., McLaughlin, M.~A., Narkevic, D.~J., \& Crawford, F. 2007, Science, 318, 777, \dodoi{10.1126/science.1147532}

\bibitem[{{Mastrano} {et~al.}(2015){Mastrano}, {Suvorov}, \& {Melatos}}]{2015MNRAS.453..522M}
{Mastrano}, A., {Suvorov}, A.~G., \& {Melatos}, A. 2015, \mnras, 453, 522, \dodoi{10.1093/mnras/stv1658}

\bibitem[{{Melatos} {et~al.}(2008){Melatos}, {Peralta}, \& {Wyithe}}]{2008ApJ...672.1103M}
{Melatos}, A., {Peralta}, C., \& {Wyithe}, J.~S.~B. 2008, \apj, 672, 1103, \dodoi{10.1086/523349}

\bibitem[{{Mereghetti} {et~al.}(2020){Mereghetti}, {Savchenko}, {Ferrigno}, {G{\"o}tz}, {Rigoselli}, {Tiengo}, {Bazzano}, {Bozzo}, {Coleiro}, {Courvoisier}, {Doyle}, {Goldwurm}, {Hanlon}, {Jourdain}, {von Kienlin}, {Lutovinov}, {Martin-Carrillo}, {Molkov}, {Natalucci}, {Onori}, {Panessa}, {Rodi}, {Rodriguez}, {S{\'a}nchez-Fern{\'a}ndez}, {Sunyaev}, \& {Ubertini}}]{2020ApJ...898L..29M}
{Mereghetti}, S., {Savchenko}, V., {Ferrigno}, C., {et~al.} 2020, \apjl, 898, L29, \dodoi{10.3847/2041-8213/aba2cf}

\bibitem[{{Morley} \& {Garcia-Pelayo}(1993)}]{1993EL.....23..185M}
{Morley}, P.~D., \& {Garcia-Pelayo}, R. 1993, EPL (Europhysics Letters), 23, 185, \dodoi{10.1209/0295-5075/23/3/005}

\bibitem[{{Peng} {et~al.}(2023){Peng}, {Wei}, \& {Wang}}]{2023ApJ...959..109P}
{Peng}, F.-K., {Wei}, J.-J., \& {Wang}, H.-Q. 2023, \apj, 959, 109, \dodoi{10.3847/1538-4357/acfcb2}

\bibitem[{{Petroff} {et~al.}(2022){Petroff}, {Hessels}, \& {Lorimer}}]{2022A&ARv..30....2P}
{Petroff}, E., {Hessels}, J.~W.~T., \& {Lorimer}, D.~R. 2022, \aapr, 30, 2, \dodoi{10.1007/s00159-022-00139-w}

\bibitem[{{Piro} {et~al.}(2021){Piro}, {Bruni}, {Troja}, {O'Connor}, {Panessa}, {Ricci}, {Zhang}, {Burgay}, {Dichiara}, {Lee}, {Lotti}, {Niu}, {Pilia}, {Possenti}, {Trudu}, {Xu}, {Zhu}, {Kutyrev}, \& {Veilleux}}]{2021A&A...656L..15P}
{Piro}, L., {Bruni}, G., {Troja}, E., {et~al.} 2021, \aap, 656, L15, \dodoi{10.1051/0004-6361/202141903}

\bibitem[{{Rajwade} {et~al.}(2020){Rajwade}, {Mickaliger}, {Stappers}, {Morello}, {Agarwal}, {Bassa}, {Breton}, {Caleb}, {Karastergiou}, {Keane}, \& {Lorimer}}]{2020rajwade}
{Rajwade}, K.~M., {Mickaliger}, M.~B., {Stappers}, B.~W., {et~al.} 2020, \mnras, 495, 3551, \dodoi{10.1093/mnras/staa1237}

\bibitem[{{Ravi} {et~al.}(2022){Ravi}, {Law}, {Li}, {Aggarwal}, {Bhardwaj}, {Burke-Spolaor}, {Connor}, {Lazio}, {Simard}, {Somalwar}, \& {Tendulkar}}]{2022MNRAS.513..982R}
{Ravi}, V., {Law}, C.~J., {Li}, D., {et~al.} 2022, \mnras, 513, 982, \dodoi{10.1093/mnras/stac465}

\bibitem[{{Ridnaia} {et~al.}(2021){Ridnaia}, {Svinkin}, {Frederiks}, {Bykov}, {Popov}, {Aptekar}, {Golenetskii}, {Lysenko}, {Tsvetkova}, {Ulanov}, \& {Cline}}]{2021NatAs...5..372R}
{Ridnaia}, A., {Svinkin}, D., {Frederiks}, D., {et~al.} 2021, Nature Astronomy, 5, 372, \dodoi{10.1038/s41550-020-01265-0}

\bibitem[{{Sang} \& {Lin}(2022)}]{2022MNRAS.510.1801S}
{Sang}, Y., \& {Lin}, H.-N. 2022, \mnras, 510, 1801, \dodoi{10.1093/mnras/stab3600}

\bibitem[{{Spitler} {et~al.}(2016){Spitler}, {Scholz}, {Hessels}, {Bogdanov}, {Brazier}, {Camilo}, {Chatterjee}, {Cordes}, {Crawford}, {Deneva}, {Ferdman}, {Freire}, {Kaspi}, {Lazarus}, {Lynch}, {Madsen}, {McLaughlin}, {Patel}, {Ransom}, {Seymour}, {Stairs}, {Stappers}, {van Leeuwen}, \& {Zhu}}]{2016Natur.531..202S}
{Spitler}, L.~G., {Scholz}, P., {Hessels}, J.~W.~T., {et~al.} 2016, \nat, 531, 202, \dodoi{10.1038/nature17168}

\bibitem[{{Tavani} {et~al.}(2021){Tavani}, {Casentini}, {Ursi}, {Verrecchia}, {Addis}, {Antonelli}, {Argan}, {Barbiellini}, {Baroncelli}, {Bernardi}, {Bianchi}, {Bulgarelli}, {Caraveo}, {Cardillo}, {Cattaneo}, {Chen}, {Costa}, {Del Monte}, {Di Cocco}, {Di Persio}, {Donnarumma}, {Evangelista}, {Feroci}, {Ferrari}, {Fioretti}, {Fuschino}, {Galli}, {Gianotti}, {Giuliani}, {Labanti}, {Lazzarotto}, {Lipari}, {Longo}, {Lucarelli}, {Magro}, {Marisaldi}, {Mereghetti}, {Morelli}, {Morselli}, {Naldi}, {Pacciani}, {Parmiggiani}, {Paoletti}, {Pellizzoni}, {Perri}, {Perotti}, {Piano}, {Picozza}, {Pilia}, {Pittori}, {Puccetti}, {Pupillo}, {Rapisarda}, {Rappoldi}, {Rubini}, {Setti}, {Soffitta}, {Trifoglio}, {Trois}, {Vercellone}, {Vittorini}, {Giommi}, \& {D'Amico}}]{2021NatAs...5..401T}
{Tavani}, M., {Casentini}, C., {Ursi}, A., {et~al.} 2021, Nature Astronomy, 5, 401, \dodoi{10.1038/s41550-020-01276-x}

\bibitem[{{Thompson} {et~al.}(2000){Thompson}, {Duncan}, {Woods}, {Kouveliotou}, {Finger}, \& {van Paradijs}}]{2000ApJ...543..340T}
{Thompson}, C., {Duncan}, R.~C., {Woods}, P.~M., {et~al.} 2000, \apj, 543, 340, \dodoi{10.1086/317072}

\bibitem[{{Tsallis}(1988)}]{1988JSP....52..479T}
{Tsallis}, C. 1988, Journal of Statistical Physics, 52, 479, \dodoi{10.1007/BF01016429}

\bibitem[{{Tsallis} {et~al.}(1998){Tsallis}, {Mendes}, \& {Plastino}}]{1998PhyA..261..534T}
{Tsallis}, C., {Mendes}, R., \& {Plastino}, A.~R. 1998, Physica A Statistical Mechanics and its Applications, 261, 534, \dodoi{10.1016/S0378-4371(98)00437-3}

\bibitem[{{Wang} {et~al.}(2023){Wang}, {Wu}, \& {Dai}}]{2023ApJ...949L..33W}
{Wang}, F.~Y., {Wu}, Q., \& {Dai}, Z.~G. 2023, \apjl, 949, L33, \dodoi{10.3847/2041-8213/acd5d2}

\bibitem[{{Wang} \& {Yu}(2017)}]{2017JCAP...03..023W}
{Wang}, F.~Y., \& {Yu}, H. 2017, \jcap, 2017, 023, \dodoi{10.1088/1475-7516/2017/03/023}

\bibitem[{{Wang} {et~al.}(2015){Wang}, {Chang}, {Wang}, \& {Lu}}]{Wang2015}
{Wang}, P., {Chang}, Z., {Wang}, H., \& {Lu}, H. 2015, European Physical Journal B, 88, 206, \dodoi{10.1140/epjb/e2015-60441-6}

\bibitem[{{Wei}(2023)}]{Wei2023}
{Wei}, J.-J. 2023, Physical Review Research, 5, 013019, \dodoi{10.1103/PhysRevResearch.5.013019}

\bibitem[{Wei {et~al.}(2021)Wei, Wu, Dai, Wang, Wang, Li, \& Zhang}]{Wei2021}
Wei, J.-J., Wu, X.-F., Dai, Z.-G., {et~al.} 2021, The Astrophysical Journal, 920, 153, \dodoi{10.3847/1538-4357/ac2604}

\bibitem[{{Wu} {et~al.}(2023){Wu}, {Zhao}, \& {Wang}}]{2023MNRAS.523.2732W}
{Wu}, Q., {Zhao}, Z.-Y., \& {Wang}, F.-Y. 2023, \mnras, 523, 2732, \dodoi{10.1093/mnras/stad1585}

\bibitem[{{Xiao} {et~al.}(2021){Xiao}, {Wang}, \& {Dai}}]{2021SCPMA..6449501X}
{Xiao}, D., {Wang}, F., \& {Dai}, Z. 2021, Science China Physics, Mechanics, and Astronomy, 64, 249501, \dodoi{10.1007/s11433-020-1661-7}

\bibitem[{Xu {et~al.}(2022)Xu, Niu, Chen, Lee, Zhu, Dong, Zhang, Jiang, Wang, Xu, Zhang, Fu, Filippenko, Peng, Zhou, Zhang, Wang, Feng, Li, Brink, Li, Lu, Yang, Caballero, Cai, Chen, Dai, Djorgovski, Esamdin, Gan, Guhathakurta, Han, Hao, Huang, Jiang, Li, Li, Li, Li, Li, Liu, Luo, Men, Niu, Peng, Qian, Song, Stern, Stockton, Sun, Wang, Wang, Wang, Wang, Wu, Xiao, Xiong, Xu, Xu, Yang, Yang, Yao, Yi, Yue, Yu, Yu, Yuan, Zhang, Zhang, Zhang, Zhao, Zheng, Zhu, \& Zou}]{Xu2022}
Xu, H., Niu, J.~R., Chen, P., {et~al.} 2022, Nature, 609, 685, \dodoi{10.1038/s41586-022-05071-8}

\bibitem[{{Younes} {et~al.}(2023){Younes}, {Baring}, {Harding}, {Enoto}, {Wadiasingh}, {Pearlman}, {Ho}, {Guillot}, {Arzoumanian}, {Borghese}, {Gendreau}, {G{\"o}{\v{g}}{\"u}{\c{s}}}, {G{\"u}ver}, {van der Horst}, {Hu}, {Jaisawal}, {Kouveliotou}, {Lin}, \& {Majid}}]{2023NatAs...7..339Y}
{Younes}, G., {Baring}, M.~G., {Harding}, A.~K., {et~al.} 2023, Nature Astronomy, 7, 339, \dodoi{10.1038/s41550-022-01865-y}

\bibitem[{{Zhang}(2020)}]{2020Natur.587...45Z}
{Zhang}, B. 2020, \nat, 587, 45, \dodoi{10.1038/s41586-020-2828-1}

\bibitem[{{Zhang}(2023)}]{2023RvMP...95c5005Z}
---. 2023, Reviews of Modern Physics, 95, 035005, \dodoi{10.1103/RevModPhys.95.035005}

\bibitem[{{Zhang} {et~al.}(2021){Zhang}, {Wang}, {Wu}, {Wang}, {Li}, {Dai}, \& {Zhang}}]{2021ApJ...920L..23Z}
{Zhang}, G.~Q., {Wang}, P., {Wu}, Q., {et~al.} 2021, \apjl, 920, L23, \dodoi{10.3847/2041-8213/ac2a3b}

\bibitem[{{Zhang} {et~al.}(2022){Zhang}, {Wang}, {Feng}, {Zhang}, {Li}, {Tsai}, {Niu}, {Luo}, {Yao}, {Zhu}, {Han}, {Lee}, {Zhou}, {Niu}, {Jiang}, {Wang}, {Zhang}, {Xu}, {Wang}, \& {Xu}}]{2022RAA....22l4002Z}
{Zhang}, Y.-K., {Wang}, P., {Feng}, Y., {et~al.} 2022, Research in Astronomy and Astrophysics, 22, 124002, \dodoi{10.1088/1674-4527/ac98f7}

\bibitem[{{Zhou} {et~al.}(2022){Zhou}, {Han}, {Zhang}, {Lee}, {Zhu}, {Li}, {Jing}, {Wang}, {Zhang}, {Jiang}, {Niu}, {Luo}, {Xu}, {Zhang}, {Wang}, {Xu}, {Wang}, {Yang}, \& {Feng}}]{2022RAA....22l4001Z}
{Zhou}, D.~J., {Han}, J.~L., {Zhang}, B., {et~al.} 2022, Research in Astronomy and Astrophysics, 22, 124001, \dodoi{10.1088/1674-4527/ac98f8}

\end{thebibliography}
\bibliographystyle{aasjournal}

\end{document}